\documentclass[a4paper,11pt]{article}
\pdfoutput=1
\usepackage{jheppub}
\usepackage[T1]{fontenc}
\usepackage{amsfonts,amssymb,epsfig,verbatim,mathbbol,amstext,amsmath,psfrag}
\usepackage{dsfont}
\usepackage{slashed}

\title{
Inverse magnetic catalysis and the Polyakov loop
}
\author[a]{Falk Bruckmann,}
\author[a]{Gergely Endr\H{o}di,}
\author[b]{and Tam\'as G.\ Kov\'acs}

\affiliation[a]{Institute for Theoretical Physics, Universit\"at Regensburg \\
                D-93040 Regensburg, Germany}
\affiliation[b]{Institute of Nuclear Research of the Hungarian Academy of
  Sciences,\\ Bem t\'er 18/c, H-4026 Debrecen,  Hungary}

\emailAdd{falk.bruckmann@physik.uni-regensburg.de}
\emailAdd{gergely.endrodi@physik.uni-regensburg.de}
\emailAdd{kgt@atomki.mta.hu}

\newcommand{\be}{\begin{equation}}
\newcommand{\ee}{\end{equation}}
\newcommand{\Tr}{\textmd{Tr}}

\newcommand{\Z}{\mathcal{Z}}
\newcommand{\D}{\mathcal{D}}

\renewcommand{\O}{\mathcal{O}}
\newcommand{\expv}[1]{\left \langle #1 \right \rangle}

\newcommand{\Seff}{S_f}

\def\p{\varphi}
\def\a{\alpha}
\def\s{s}

\abstract{ We study the physical mechanism of how an external magnetic field
  influences the QCD quark condensate. Two competing mechanisms are identified,
  both relying on the interaction between the magnetic field and the low quark
  modes. While the coupling to valence quarks enhances the condensate, 
  the interaction with sea quarks suppresses it in the transition region. The latter ``sea effect''
  acts by ordering the Polyakov loop and, thereby, reduces the number of small
  Dirac eigenmodes and the condensate. It is most effective around the
  transition temperature, where the Polyakov loop effective potential is flat and
  a small correction to it by the magnetic field can have a significant
  effect. Around the critical temperature, the sea suppression overwhelms the
  valence enhancement, resulting in a net suppression of the condensate, named
  inverse magnetic catalysis. We support this physical picture by lattice
  simulations including continuum extrapolated results on the Polyakov loop as
  a function of temperature and magnetic field.
We argue that taking into account the increase in the Polyakov loop and its interaction 
with the low-lying modes is essential to obtain the full physical picture, and should be 
incorporated in effective models for the description of QCD in magnetic fields 
in the transition region.
}

\begin{document} 
 \maketitle
\flushbottom

\section{Introduction}
   \label{sec:I}

It is by now well established that strongly interacting matter has at least two
distinct forms of existence: the hadronic and the quark-gluon plasma
phase. They are separated by a crossover~\cite{Aoki:2006we}, occurring at a
transition temperature $T_c$ of about 150 MeV~\cite{Aoki:2006br,Borsanyi:2010bp,Bazavov:2011nk}. Understanding the
thermodynamics of strongly interacting matter around this transition is essential for the 
description of systems including the early universe, the interior of neutron stars or heavy-ion experiments. 
In heavy-ion collisions, for instance, the colliding beams can create extremely high magnetic fields 
at the collision center, which can
influence the thermodynamic properties of the
system~\cite{Skokov:2009qp,Voronyuk:2011jd,Bzdak:2011yy,Deng:2012pc}. 
Similarly, the external magnetic field is expected to play an important role for very dense neutron stars (see, e.g., ref.~\cite{Duncan:1992hi}) and for the evolution of the early universe (see, e.g., ref.~\cite{Vachaspati:1991nm}).

A particularly pronounced effect of the external field $B$ is its influence
on chiral symmetry breaking. At zero temperature, both low energy effective
theories (see,
e.g., refs.~\cite{Klevansky:1989vi,Schramm:1991ex,Klimenko:1992ch,Gusynin:1995nb,Shushpanov:1997sf}, and ref.~\cite{Shovkovy:2012zn} for a review) and early lattice
simulations predicted the magnetic field to enhance the quark
condensate. Although these first lattice simulations used the quenched approximation~\cite{Buividovich:2008wf}
and larger-than-physical quark masses~\cite{D'Elia:2010nq,D'Elia:2011zu}, the
effect was later also confirmed by lattice simulations at physical quark
masses in the continuum limit~\cite{Bali:2011qj,Bali:2011uf,Bali:2012zg}.  We
can thus safely conclude that at zero temperature, the external magnetic field
enhances the condensate. This phenomenon is called {\it magnetic catalysis}.
We remark that the zero temperature magnetic catalysis of the condensate can also be derived from the positivity 
of the scalar QED $\beta$-function, thereby showing its universal nature~\cite{Endrodi:2013cs}.

At finite temperature, and especially around $T_c$ -- in the temperature range, which is most
important for heavy ion collisons -- the situation is not so clear. Here, most
of the low energy models also predict that magnetic
catalysis takes place, and, furthermore, $T_c$ is shifted to higher temperatures 
(for exceptions see, e.g., refs.~\cite{Agasian:2008tb,Fraga:2012fs}, and ref.~\cite{Ayala:2012dk} for a bosonic system with charged pion condensates).
This behavior was also observed for larger-than-physical quark masses on
coarse lattices~\cite{D'Elia:2010nq,Ilgenfritz:2012fw}. In contrast,
simulations on finer grids and with physical light quark masses indicated that
around the crossover temperature exactly the opposite happens: the magnetic
field suppresses the quark condensate and shifts the crossover temperature
downwards~\cite{Bali:2011qj,Bali:2011uf,Bali:2012zg}. The discrepancy between
the different lattice results suggests that the use of fine grids and light
enough quarks~\cite{Bali:2011qj} might be essential for correctly describing
the effect.
Moreover, the remarkable similarity between the reaction of gluonic and fermionic observables to the magnetic field~\cite{Bali:2013esa} implies that the indirect effect of $B$ on the gauge degrees of freedom -- exerted through the electrically charged quarks -- must be a crucial ingredient for the description of the transition region.

In the present paper, we study the physical mechanism behind the suppression
of the condensate by the magnetic field in the transition region -- an effect
we refer to as {\it inverse magnetic catalysis}\footnote{We note that the
  expression `inverse catalysis' is also used for the reduction of the
  transition temperature with growing $B$ at nonzero chemical
  potential~\cite{Preis:2010cq}, but presumably has a completely different
  physical origin.}.
We will show that the magnetic field influences chiral symmetry breaking
through two different mechanisms: one enhances and the other suppresses the
condensate.
Let us first describe the mechanism that enhances the condensate and is also
responsible for catalysis at zero temperature. According to the Banks-Casher
relation~\cite{Banks:1979yr}, the quark condensate is proportional to the
spectral density of the Dirac operator around zero. The Dirac operator
explicitly contains the magnetic field and, thus, its spectral density in a
fixed gauge background depends on $B$. As we will show, the magnetic field
enhances the spectral density around zero and, therefore, also enhances the
quark condensate. In the free case as well as most of the model calculations
this enhancement is linked to the degeneracy of eigenvalues being proportional
to the magnetic flux. This is purely a ``valence'' effect that can already be
observed in the quenched approximation, where the back-reaction of the quarks
on the gauge field is ignored.
We can view this as the basic mechanism explaining magnetic catalysis.

Besides the explicit dependence of the Dirac operator on $B$, the magnetic field 
also enters the
quark action, and influences the probability of different gauge backgrounds in
the path integral. We find that this dependence leads to the suppression of the condensate in the transition region. This is a ``sea'' effect, since its origin is the
different sampling of the gauge fields due to the quark determinant. 
The distinction between valence
and sea effects was initially introduced in ref.~\cite{D'Elia:2011zu} with a slightly different terminology (the ``sea'' was called
``dynamical''). 

Although the two mechanisms are of very different nature, we will show that both of them 
originate from the reaction of the low-lying Dirac modes to the magnetic field. However, the magnetic field ``in the sea'' has exactly the opposite effect in the transition region as the ``valence'' mechanism.
Thus, the relative
strength of the two effects will determine whether the QCD condensate
undergoes magnetic catalysis or, rather, inverse magnetic catalysis\footnote{
  We note that possible explanations of the lattice
  results~\cite{Bali:2011qj,Bali:2011uf,Bali:2012zg} regarding inverse
  magnetic catalysis were given recently in
  Refs.~\cite{Fukushima:2012kc,Kojo:2012js} relying on the dimensional
  reduction of the lowest Landau level dynamics for large
  magnetic fields, $eB > \Lambda_{\rm QCD}^2$, plus screening and neutral meson effects.
  We remark,
  that the lattice data indicate
  that inverse magnetic catalysis is related to the transition taking place, since for even
  higher temperatures $T\gg T_c$ the conventional magnetic catalysis again
  prevails~\cite{Bali:2012zg}.}. As it turns out, the most important control parameter that influences the competition of 
the valence and sea effects, is the quark mass $m$. In particular, the two effects are 
expected to have a drastically different 
$m$-dependence. On the one hand, the valence effect is already seen in
the quenched approximation, and is expected to depend on $m$ only mildly. 
On the other hand, the sea effect is completely absent in the
quenched approximation -- which technically corresponds to infinite quark masses --
and is expected to increase as the quarks become lighter. This shows that the
quark masses have to be tuned to their physical values in order to have a
proper description of magnetic (inverse) catalysis.

Besides identifying the sea mechanism as the one responsible for inverse magnetic catalysis, it is instructive to look at how the magnetic field in the determinant changes
the typical gauge configurations contributing to the path
integral. Understanding this is clearly essential for the incorporation of the sea effect
into effective models describing QCD. As we will show, one of 
the most striking influences of the magnetic field on the gauge configurations
is to drive up the expectation value of the Polyakov loop\footnote{That the Polyakov loop plays a role for the catalysis mechanism in the transition region has been advocated in Refs.~\cite{Fukushima:2012xw,Andersen:2012jf}.
 }. 
This can be understood by noting that the Polyakov loop has a strong
influence on the lowest part of the Dirac spectrum. Namely, the ordering of
the Polyakov loop shifts up the small Dirac eigenvalues and thus suppresses
the quark condensate. 
However, the Polyakov loop is
also influenced by the effective action originating from the gauge action and
the fermion action at zero magnetic field. The change in this effective action
caused by the magnetic field is usually relatively small, and cannot be
decisive. The only exception to this is around the transition temperature, where
the Polyakov loop effective action is flat, and a small contribution coming
from the magnetic field can have a large ordering effect on the Polyakov loop. This, in turn, suppresses low Dirac eigenmodes and the condensate. 
We conjecture that this is the physical mechanism behind
inverse magnetic catalysis seen around the transition temperature.

The rest of the paper is devoted to a demonstration of the physical picture we
just described using numerical lattice simulations. The results we present are
partly based on large scale simulations already reported in
refs.~\cite{Bali:2011qj,Bali:2012zg}, and partly on new simulations involving
small lattices, where a full diagonalization of the Dirac operator was
feasible.  In each case, we use $1+1+1$ flavors of stout smeared staggered
quarks with physical quark masses and the tree-level improved Symanzik gauge
action. Details of the simulation setup and the configurations can be found in
refs.~\cite{Aoki:2005vt,Borsanyi:2010cj,Bali:2011qj}.

The paper is organized as follows. In the next section, we define the valence
and sea contribution to the quark condensate, and measure them for several
magnetic fields at temperatures below and close to the transition. In
Sec.~\ref{sec:CttPl}, we discuss the effects of the magnetic field on the
Polyakov loop, and Sec.~\ref{sec:C} contains our conclusions.  In the appendices we calculate the interplay of magnetic field, Polyakov loop and mass in the free case, and discuss the renormalization properties of the sea and valence condensates and the Polyakov loop.

\section{Sea and valence quark effects}
  \label{sec:Savqe}

As we already noted, the most important notion, that connects all different
aspects of the problem, is the spectrum of the quark Dirac operator. The Dirac
spectrum is affected by the magnetic field for two different reasons.  On the
one hand, the magnetic field explicitly appears in the Dirac operator, and
directly influences its spectrum in any fixed gauge background. On the other
hand, through the quark determinant, $B$ also affects the probability measure
of how the gauge field configurations are sampled in the path
integral. Throughout this paper, we will call the first one the ``valence''
effect, and the second one the ``sea'' effect. In this section, we introduce
our notation, and discuss these effects in detail.

In physical terms, the most important measurable property of the Dirac
spectrum is the spectral density $\rho(\lambda)$ around zero, which is
proportional to the chiral condensate due to the Banks-Casher
relation~\cite{Banks:1979yr},
\be
\lim_{\lambda\to0} \lim_{V\to\infty} \rho(\lambda)\cdot \pi = \lim_{m\to0}
\lim_{V\to\infty} \bar\psi\psi, 
\quad\quad\quad
\bar\psi\psi = \frac{T}{V} \frac{\partial \log\Z}{\partial m}.
\label{eq:pbpdef}
\ee
Here the condensate is defined in terms of the partition function, which is
given by the path integral over gauge configurations $U$ as
\be
\Z(B) = \int \D U\, e^{-S_g} \det ( \slashed{D}(B)+m),
\label{eq:partfunc}
\ee
where $S_g$ denotes the gauge action. For the sake of clarity, here we consider only one fermion flavor
with charge $q$ and mass $m$, and suppress factors of $1/4$, which appear due
to the rooting procedure for staggered quarks. The temperature and the three-volume are
given as $T=(aN_t)^{-1}$ and $V=(aN_s)^3$ with $N_s$ ($N_t$) the number of
lattice sites in the spatial (temporal) direction, and $a$ the lattice
spacing.  We remark that since the magnetic
field couples only to the electric charges of quarks, it enters exclusively in the
combination $qB$.

Expanding the derivative in Eq.~(\ref{eq:pbpdef}), the condensate is obtained as
\be
\bar\psi\psi(B) = \frac{1}{\Z(B)}\int \D U\, e^{-S_g} \det
(\slashed{D}(B)+m)\, \Tr(\slashed{D}(B)+m)^{-1}, 
\label{eq:pbptot}
\ee
showing that the magnetic field indeed appears both in the determinant and in
the operator itself. To separate these dependences, we define the valence and
sea condensates as
\be
\begin{split}
\bar\psi\psi^{\rm val}(B) &= \frac{1}{\Z(0)}\int \D U\, e^{-S_g} \det
(\slashed{D}(0)+m)\, \Tr(\slashed{D}(B)+m)^{-1}, \\ 
\bar\psi\psi^{\rm sea}(B) &= \frac{1}{\Z(B)}\int \D U\, e^{-S_g} \det
(\slashed{D}(B)+m)\, \Tr(\slashed{D}(0)+m)^{-1}. 
\end{split}
\label{eq:pbpvalsea}
\ee
We note that valence condensates can be used to define dressed Wilson loops~\cite{Bruckmann:2011zx}, 
which are directly related to the QCD string tension in the large mass limit.

Any physically consistent theory has to have the same valence and sea fermion
content. Thus, at first sight it does not seem possible to separate the
valence and sea effects of the magnetic field in a well-defined theory. 
One can, nevertheless, exactly reproduce the condensates in Eq.~(\ref{eq:pbpvalsea}) by
alluding to techniques from partially quenched QCD
\cite{Morel:1987xk,Sharpe:2006pu,Verbaarschot:2000dy} (and quenched disorder
\cite{Brezin:1984,Efetov:1997fw}). Using commuting spin $1/2$-fields (so-called ghost
quarks), one can generate inverse Dirac determinants in the functional integral. 
With adjusted charges and masses, these inverse determinants can cancel `unwanted' determinants in the path integral, arriving at the valence and sea condensates of Eq.~(\ref{eq:pbpvalsea}).
From a different point of view, one can also directly obtain the sea condensate in 
a theory with an electrically charged and a neutral fermion flavor,
by looking at the condensate of the neutral fermion in the presence of the magnetic field.
Since $B$ appears in the determinant of the charged flavor, but not in the neutral propagator, 
this indeed isolates the sea effect. 
Even though in QCD all fermion species are electrically charged, on the technical 
level of the lattice theory, the valence and sea effects are naturally separated.
We will use a similar argument in App.~\ref{app:renormseaval} to discuss the renormalization of the sea and valence condensates. 

To discuss the effect of the external magnetic field, we are interested in the
change of the condensates due to a nonzero $B$. This change is given by the
difference 
\be \Delta \Sigma(B) =
\frac{2m}{M_\pi^2 F^2} \left[
    \bar\psi\psi(B)- \bar\psi\psi(0) \right].
\label{eq:pbpren}
\ee
This combination is particularly useful, because both additive and
multiplicative divergences cancel in it. Based on the Gell-Mann-Oakes-Renner relation, the normalization is chosen
such that $\Delta\Sigma$ is measured in units of the condensate at zero
magnetic field and zero temperature~\cite{Bali:2012zg}.  We define
$\Delta\Sigma^{\rm val}$ and $\Delta\Sigma^{\rm sea}$ in a similar manner from
Eq.~(\ref{eq:pbpvalsea}).  At $B=0$ the three types of condensate are obviously
equal. Furthermore, for small magnetic fields (assuming analyticity in $qB$),
the two contributions appear additively in the total
condensate~\cite{D'Elia:2011zu},
\be
\Delta\Sigma(B) \simeq \Delta\Sigma^{\rm val}(B) + \Delta\Sigma^{\rm sea}(B),
\label{eq:seavaladd}
\ee
showing that this separation indeed makes sense, at least for small magnetic
fields. In App.~\ref{app:renormseaval} we show that $\Delta\Sigma^{\rm val}$ and $\Delta\Sigma^{\rm sea}$ are both properly renormalized.

\begin{figure}[t!]
\centering 
\includegraphics[width=.5\textwidth]{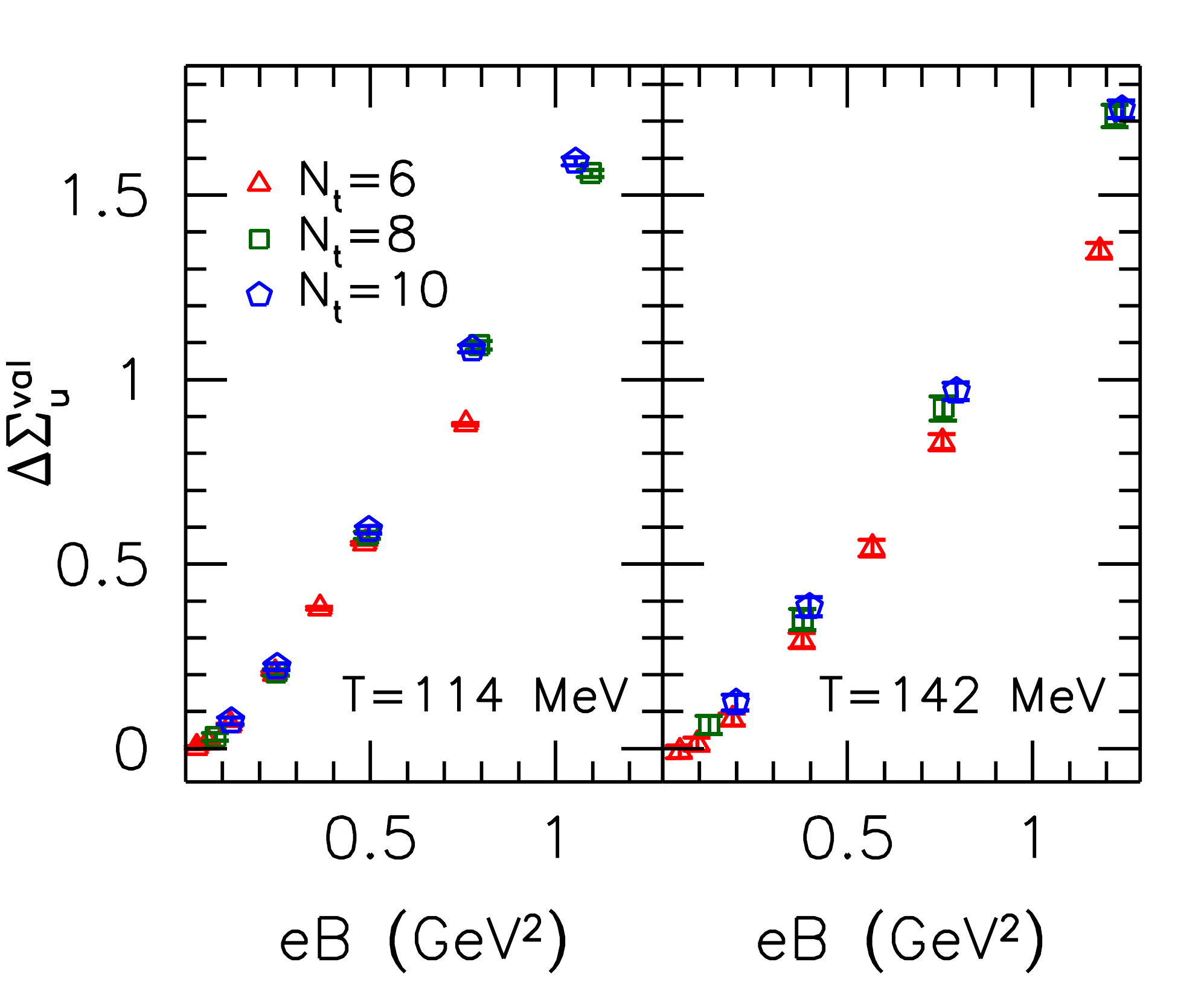}
\hspace*{-.2cm}
\includegraphics[width=.5\textwidth]{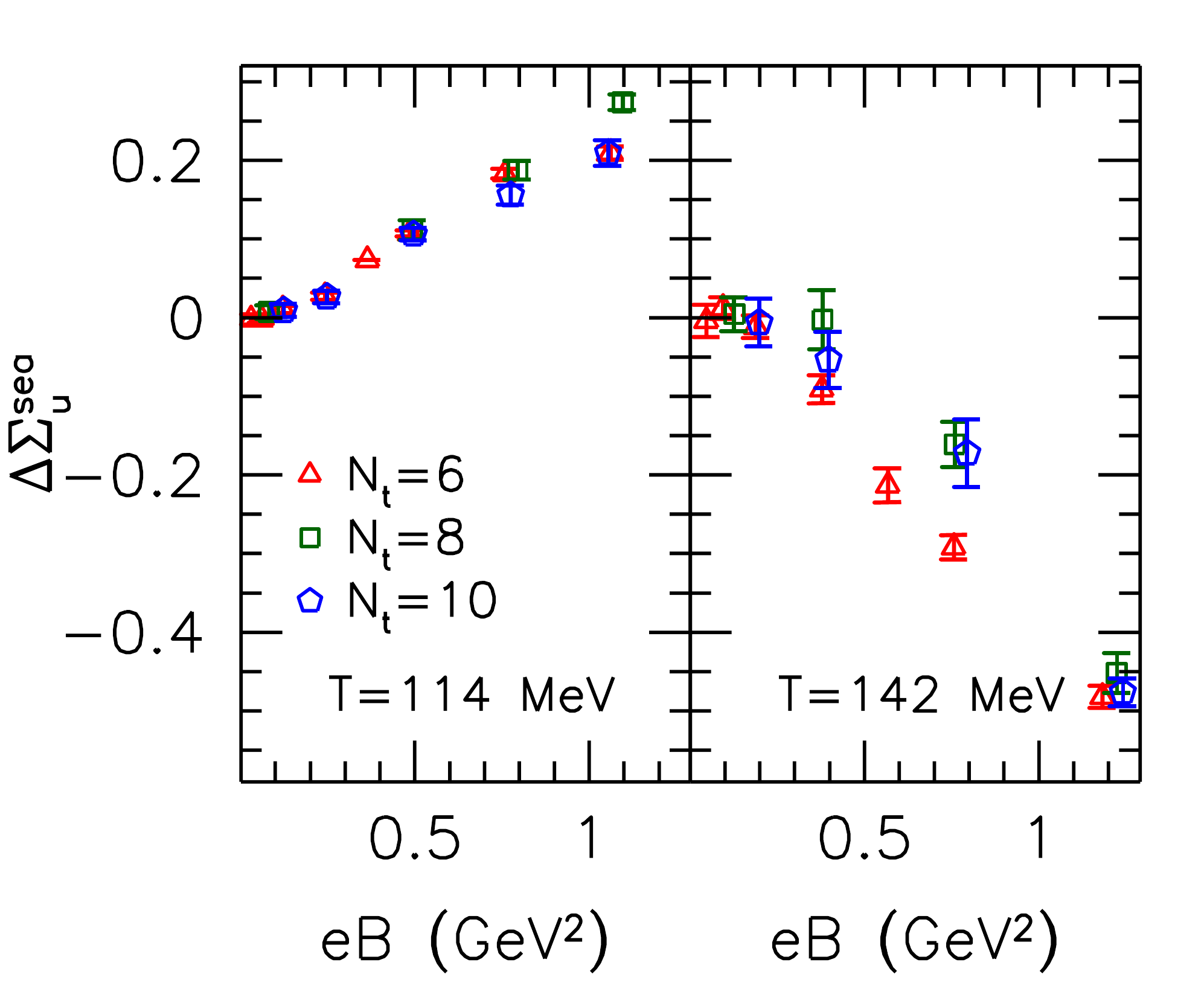}
\hspace*{-.2cm}
\caption{\label{fig:valence_sea} The valence (left panel) and sea (right
  panel) contributions to the up quark condensate as a function of the
  magnetic field, calculated at two different temperatures. The two
  temperatures are chosen to be well below and around $T_c$. The different
  symbols correspond to three different lattice spacings (decreasing as
  $N_t$ grows).}
\end{figure}

In Fig.~\ref{fig:valence_sea}, we show how the valence condensate
$\Delta\Sigma^{\rm val}$ (left panels) and the sea condensate
$\Delta\Sigma^{\rm sea}$ (right panels) for the up quark depend on the
magnetic field, at two different temperatures.  The two temperatures were
chosen to be well below and just below the transition temperature. Clearly,
the magnetic field in the valence Dirac operator enhances the condensate at
both temperatures. In contrast, the sea effect enhances the condensate only
well below $T_c$, whereas around $T_c$ it suppresses it. Eventually -- around $T=160\textmd{ MeV}$ -- the sea contribution becomes the dominant one, resulting in a
decrease in the total condensate and thus inverse magnetic
catalysis~\cite{Bali:2012zg}. Fig.~\ref{fig:valence_sea} contains results for
three different lattice spacings, showing that the effect persists in the
continuum limit as well.  This is in sharp contrast to the findings of
ref.~\cite{D'Elia:2010nq}, where an enhancement was found at all temperatures, for larger-than-physical quark masses.
In Secs.~\ref{sec:thevalence} and~\ref{sec:thesea}, we discuss how the
underlying mechanism responsible for the valence and the sea effects can be
understood, based on eigenvalues of the Dirac operator at nonzero magnetic
fields.

\subsection{The valence effect}
\label{sec:thevalence}

The valence effect can be easily understood by inspecting how the low part of
the spectral density $\rho(\lambda)$ of the Dirac operator depends on the
magnetic field in any gauge field background. To be specific, we use a set of
gauge field backgrounds generated at zero magnetic field, but the qualitative
picture is the same in any reasonable ensemble of gauge fields.  

\begin{figure}[t!]
\centering 
\includegraphics[width=.5\textwidth]{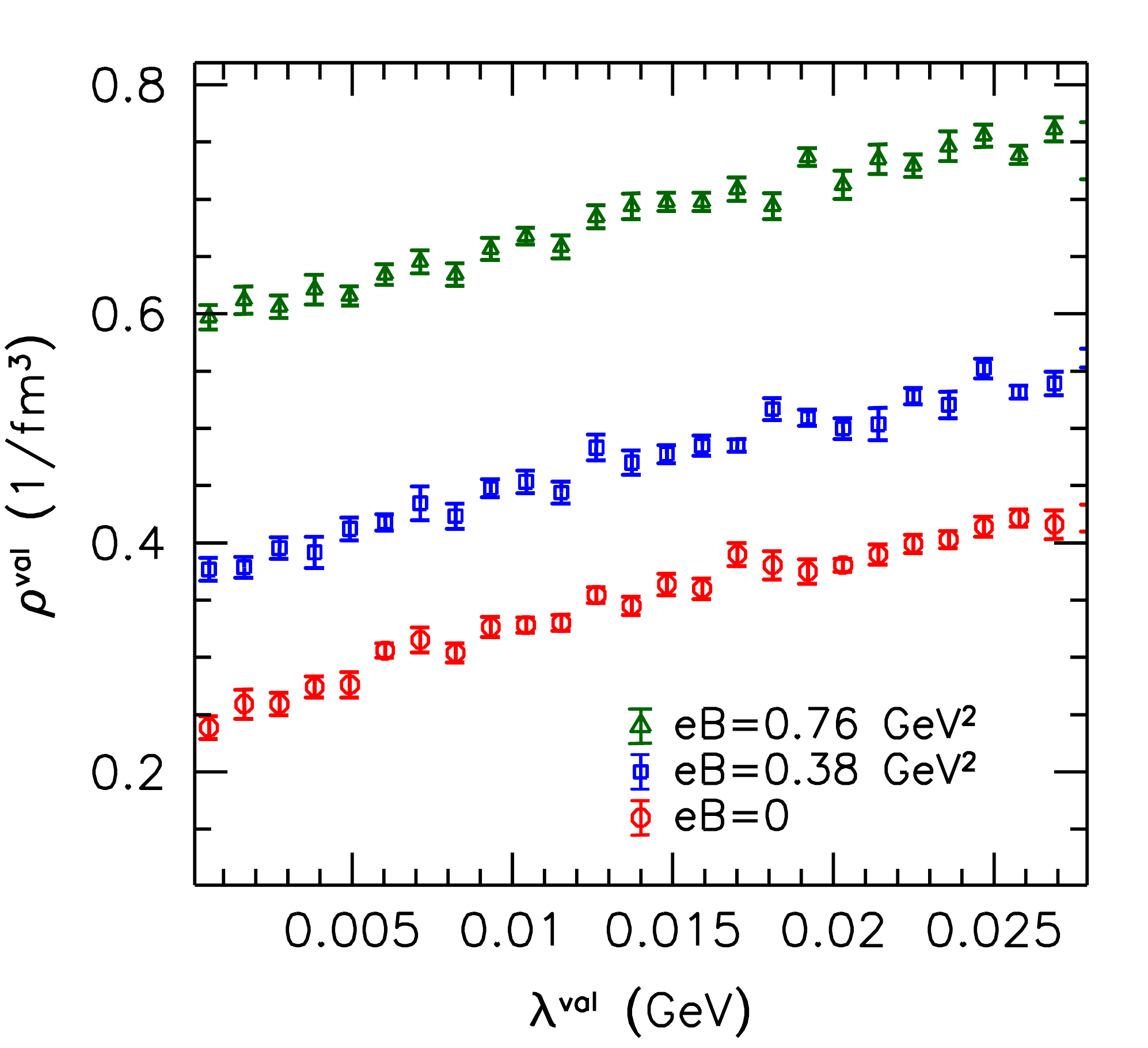}
\caption{\label{fig:specdens} The spectral density of the Dirac operator
  around zero, computed at three different values of the magnetic field. In
  all three cases, the gauge configurations were generated without the magnetic
  field in the quark determinant.}
\end{figure}

In Fig.~\ref{fig:specdens}, we plot the spectral density of the Dirac operator
for three different values of the (valence) magnetic field, as measured on
$N_t=6$ lattices, generated at $B=0$ and $T=142\textmd{ MeV}$.  This gauge
field ensemble corresponds to the \mbox{$N_t=6,\, T=142$~MeV} data for the
valence condensate in Fig.~\ref{fig:valence_sea}. One can clearly see the increase of the spectral density and, thus, of the valence condensate with the magnetic field. We remark that the same
behavior is reproduced for any gauge background, independent of the
temperature and magnetic field, which was used for the generation of the
configuration. In other words, this means that the change in the valence
condensate $\Delta\Sigma^{\rm val}$ is always positive.  We note that
a similar proliferation of low Dirac eigenmodes already occurs in the free
theory, see the discussion in App.~\ref{app:free}. 
Moreover, a remarkable 
feature of the free spectrum on the lattice is that the eigenvalue
pattern as a function of the magnetic field is similar to the so-called Hofstadter 
butterfly, the energy levels of Bloch electrons in a magnetic field~\cite{Hofstadter:1976zz}.

\subsection{The sea effect}
\label{sec:thesea}

The sea effect arises, because the magnetic field in the quark determinant
changes the relative weight of the gauge configurations, and is therefore equivalent to a reweighting in $B$. In general, reweighting is a technique that uses configurations generated at a given (starting) point of the parameter space, and assigns a new weight to each configuration, in a fashion that the resulting ensemble describes the system at a new (target) point of the parameter space. Thus, the expectation value of an arbitrary observable $\mathcal{O}$ at the target point is obtained in terms of measurements on the configurations generated at the starting point. Here, we will consider the $B=0$ system as the starting point, and the $B>0$ system as the target ensemble. For this case, the difference of weights equals the ratio of quark determinants at $B$ and at $B=0$, and the exact rewriting of the expectation value at $B$ reads
\be
\begin{split}
\expv{\O}_{B} &= \frac{\Z(0)}{\Z(B)}\cdot\frac{1}{\Z(0)} \int \D U e^{-S_g} \det \left(\slashed{D}(0)+m\right) \frac{\det(\slashed{D}(B)+m)}{\det(\slashed{D}(0)+m)} \,\O \\
&= \expv{e^{-\Delta S_f(B)}\,\O}_0\Big/ \expv{e^{-\Delta S_f(B)}}_0,
\end{split}
\ee
where the subscript of the expectation value indicates the value of the magnetic field, at which the ensemble is generated. Here, we defined the change in the fermionic action due to the magnetic field,
\be 
     -\Delta \Seff(B) = \log \det(\slashed{D}(B)+m) - 
                             \log \det(\slashed{D}(0)+m).
\label{eq:ldet}
\ee

\begin{figure}[ht!]
\centering
\includegraphics[width=.5\textwidth]{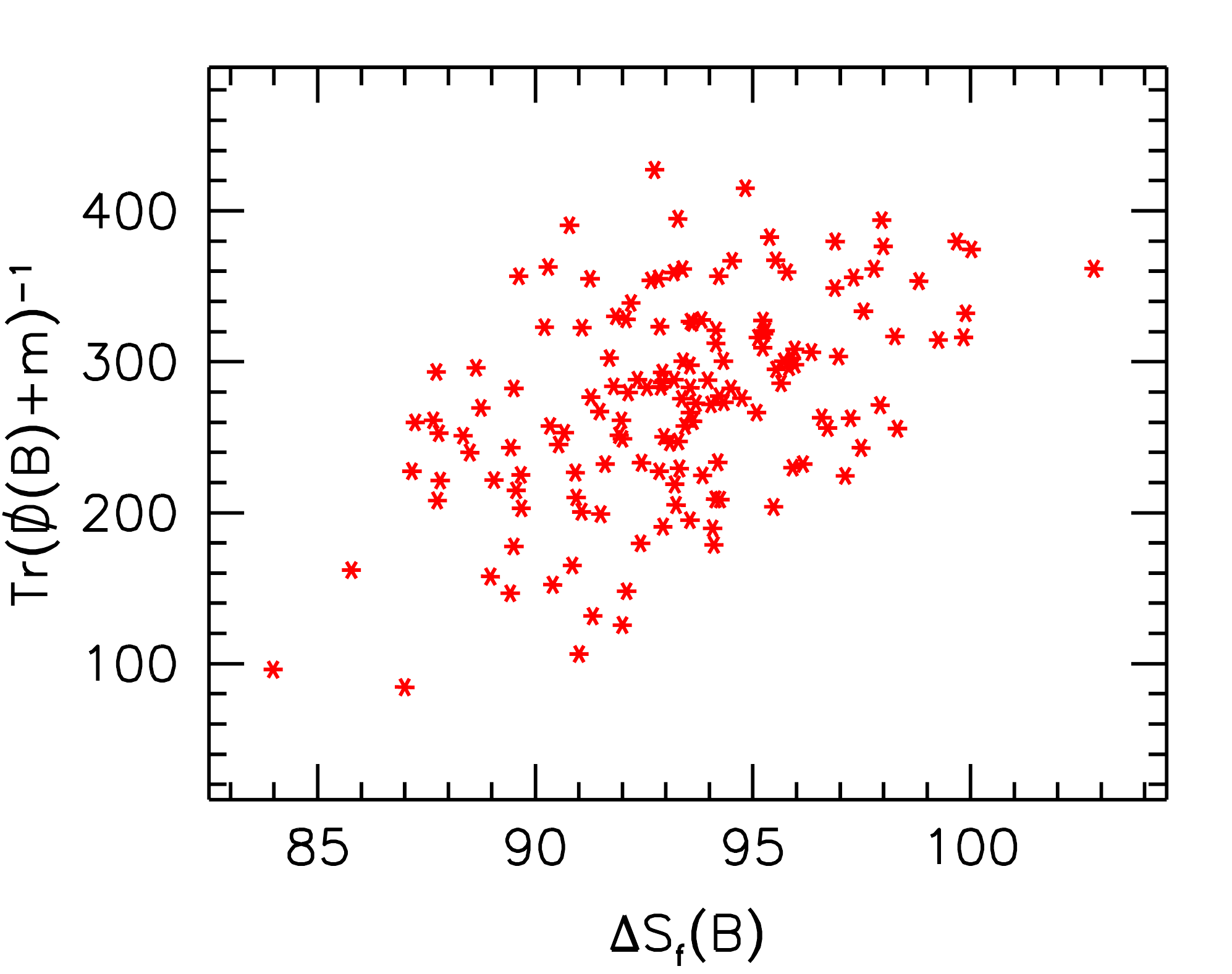}
\caption{\label{fig:scatter1} Scatter plot of the down condensate versus the change in
  the fermionic action due to the magnetic field, Eq.~(\protect\ref{eq:ldet}), at $eB\approx0.5 \textmd{ GeV}^2$. 
Each point represents a $10^3 \times 4$ gauge configuration. }
\end{figure}

Let us now apply this machinery for the case of the condensates. The valence condensate of Eq.~(\ref{eq:pbpvalsea}) equals the $B=0$ expectation value,
\be
\bar\psi\psi^{\rm val}(B)= \left\langle \Tr(\slashed{D}(B)+m)^{-1}\right\rangle_{0},
\label{eq:valence2}
\ee
whereas the full condensate is obtained in the reweighting picture as
\be
\bar\psi\psi(B) = \left\langle e^{-\Delta \Seff(B)}\,\Tr(\slashed{D}(B)+m)^{-1}\right\rangle_{0}\Big/
\left\langle e^{-\Delta \Seff(B)}\right\rangle_{0}.
\label{eq:reweighting}
\ee
Now we are in the position to discuss the sea effect, i.e.\ the difference
between the full and valence condensates. The full condensate is written in Eq.~(\ref{eq:reweighting}) as the product of the trace of the propagator at nonzero $B$ and the factor $\exp(-\Delta \Seff(B))$, both of which depend on the configuration. 
Therefore, we have to study the correlation between 
the fermion action difference and the value of the condensate at nonzero $B$. 
To this end, in Fig.~\ref{fig:scatter1} we
show a scatter plot of the down quark condensate versus $\Delta \Seff(B)$ 
for an ensemble of 150 $10^3\times4$ lattices generated at $B=0$,
and at inverse gauge coupling $\beta=3.35$, corresponding to a temperature around
$T_c$. On this gauge ensemble, we computed the condensate by inserting
$eB\approx 0.5 \textmd{ GeV}^2$ in the Dirac operator. Both the action and the
condensate were calculated in the spectral representation, by explicitly
diagonalizing the quark Dirac operator.

In Fig.~\ref{fig:scatter1}, each point represents a gauge configuration. The
ordinate is the condensate and the abscissa is the action difference on that
particular gauge configuration. Since the configurations were generated at
$B=0$, a simple averaging of the ordinates in Fig.~\ref{fig:scatter1} would
give the ``valence'' condensate, in accordance with Eq.~(\ref{eq:valence2}). To obtain the full physical condensate,
including the valence and sea effects, one needs to weight each ordinate with
the Boltzmann factor $\exp(-\Delta \Seff)$. This means that the
contribution of points corresponding to larger abscissas are exponentially
suppressed. A clear tendency can be seen in the figure, showing that larger values of
the condensate generally correspond to larger actions, and are therefore
suppressed by the quark determinant. This is the ``sea'' effect.

\begin{figure}[ht!]
\centering
\includegraphics[width=.5\textwidth]{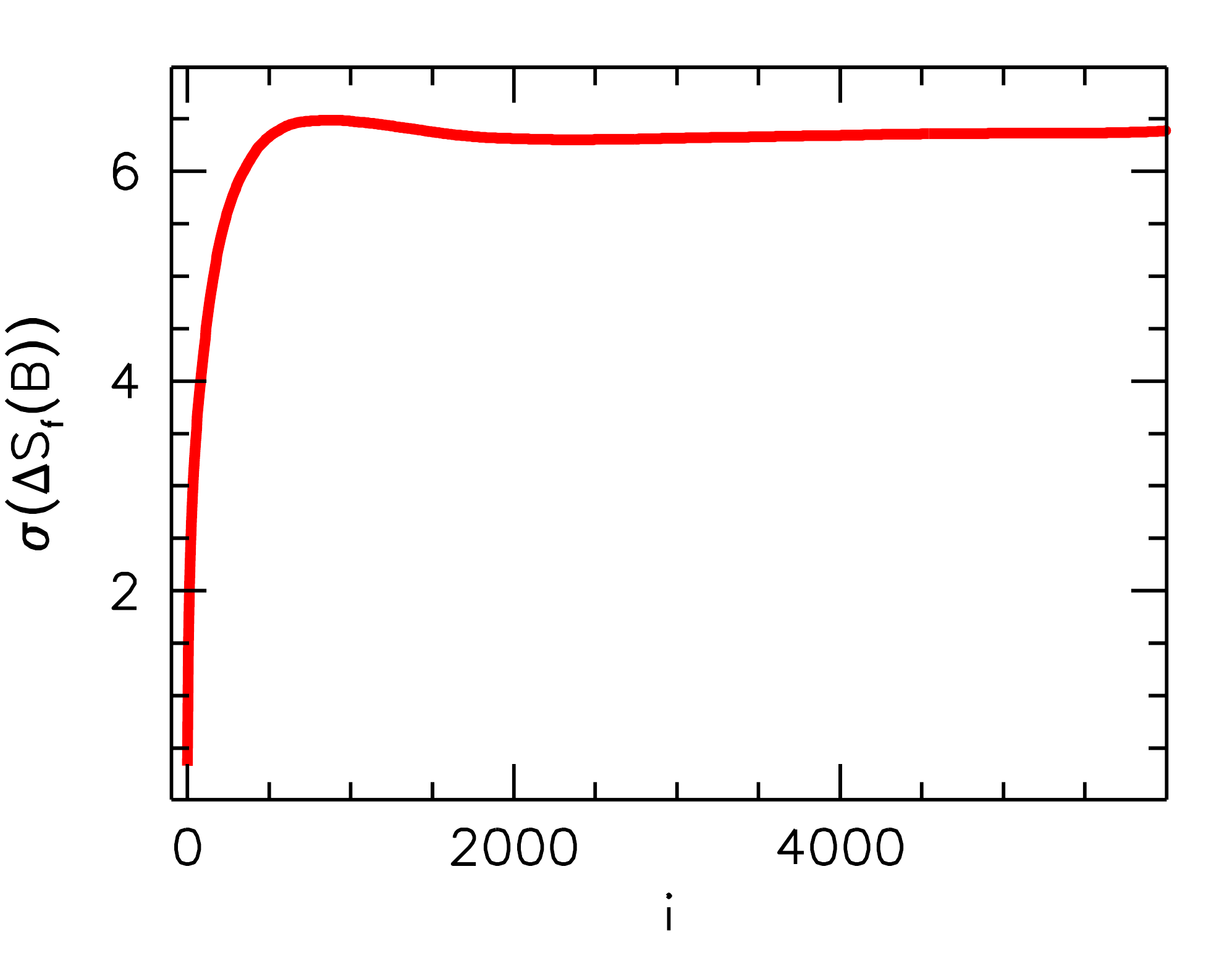}
\caption{\label{fig:reweightfac}The variation of the change in the effective
  action $\Delta \Seff$ (at $eB\approx0.5 \textmd{ GeV}^2$), calculated
  using the lowest $i$ eigenmodes, as a function of $i$. The total number of eigenvalues on this lattice is 6000.}
\end{figure}

The reweighting in the magnetic field can substantially influence the
condensate, because the reweighting factors $\Delta \Seff(B)$ fluctuate
sufficiently strongly. In Fig.~\ref{fig:scatter1}, for example, a distance of 7
on the abscissa amounts to a change in the path integral weight by three
orders of magnitude. It is known that the light quark condensate is dominated
by low Dirac modes. We now show that the fluctuations of the change in the
quark action also have their origin in the low spectral region of the Dirac
operator. 
In Fig.~\ref{fig:reweightfac}, we plot how the variance
$\sigma(x)=\sqrt{\expv{x^2}-\expv{x}^2}$ of the change in the fermion action
$x=\Delta \Seff(B)$ builds up starting from the low-end of the Dirac
spectrum. We approximate the determinants by the product of the lowest $i$
eigenvalues, and plot the variance of this ``approximated'' $\Delta \Seff(B)$ 
as a function of $i$, on the same gauge ensemble that was used
above. Looking at the figure, the dominance of the low Dirac modes in this
quantity is obvious. 

We emphasize that these reweighting factors encode all information\footnote{ On the technical level, due to finite statistics, the applicability of reweighting is limited to starting and target systems whose important configurations are not too different. This potential overlap problem is not relevant here, since we use reweighting to unravel a mechanism and the results can always be checked by direct simulations of the target system (nonzero $B$).} about
how the system evolves and how different observables change as the magnetic
field grows. The magnetic field can have an influence on an observable only if
two conditions are met: the reweighting factors have a non-zero variance and
they are correlated with the given physical quantity. Since the variance of
the reweighting factors comes mostly from the low Dirac modes, they are
expected to be correlated with any other quantity that is sensitive to the low
Dirac eigenvalues. In this context, the sea effect of the magnetic field on the
condensate is naturally expected.

\section{Connection to the Polyakov loop}
  \label{sec:CttPl}

We have seen that the quark determinant tends to suppress gauge configurations
with larger values of the quark condensate. To understand the reason behind this, 
it is desirable to identify those gauge field degrees of
freedom that play the most important role in this mechanism. Such a
description also facilitates the incorporation of this effect into
low-energy effective models. 

Around the transition temperature, the quantity depending most sensitively on
the control parameters is the order parameter. In the case of QCD, the
approximate gauge field order parameter around $T_c$ is the (averaged real part of the traced) Polyakov loop,
\be 
   P = \frac{1}{V} \expv{\sum_{\mathbf{x}} \mbox{Re} \Tr \prod_{t=0}^{N_t-1}
      U_4(\mathbf{x},t)}.  
\ee
In the valence/sea language the Polyakov loop is a purely ``sea'' observable, since the expression to be averaged over does not depend explicitly on $B$ (this applies to every gluonic observable).
    
\begin{figure}[ht!]
\centering
\includegraphics[width=.5\textwidth]{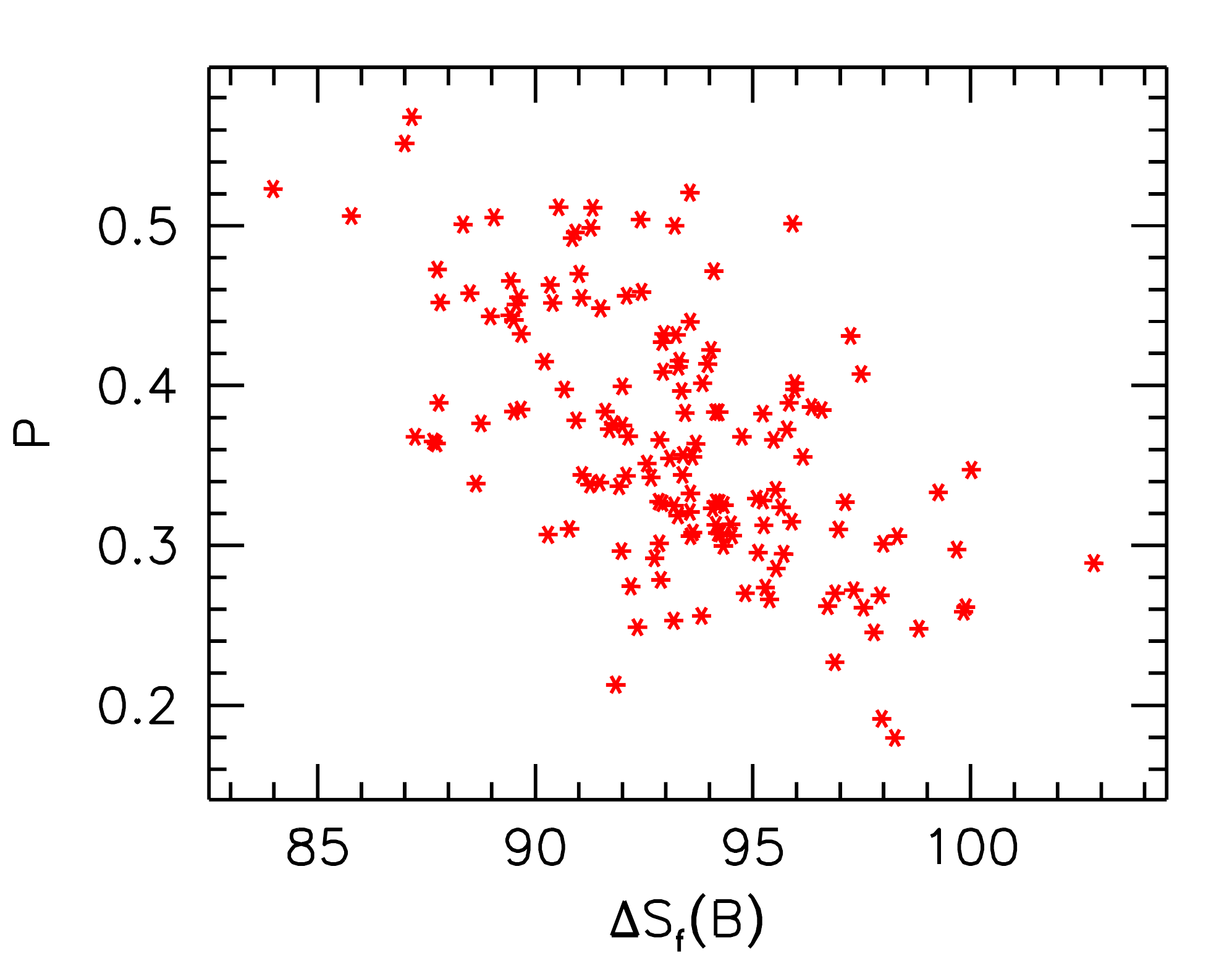}
\caption{\label{fig:scatter2} Scatter plot of the Polyakov loop versus the change in
  the fermionic action due to the magnetic field at $eB\approx0.5 \textmd{ GeV}^2$. 
Each point represents a $10^3 \times 4$ gauge configuration. }
\end{figure}

To see how the Polyakov loop reacts to the magnetic field, in Fig.~\ref{fig:scatter2},
 we show a scatter plot of $P$ versus the change $\Delta \Seff(B)$ in
the fermionic action due to the magnetic field. This plot is similar
in spirit to Fig.~\ref{fig:scatter1}, and is based on the same gauge
configurations. The two plots also have a qualitatively similar appearance,
but in the case of the Polyakov loop, the correlation is opposite. 
Here a simple average over the ordinates would give the $B=0$ Polyakov loop, whereas a weighted average yields the Polyakov loop at nonzero $B$. 
The fermionic action, and hence the magnetic field, suppresses small values and favors large values of the Polyakov loop.
Below we will argue, that large Polyakov loops point to smaller condensates.

It is also not hard to understand why this effect is so pronounced around the
transition temperature. Since there is a crossover at $T_c$, the effective
potential for the Polyakov loop is flat there. As a result, even a small
contribution coming from the magnetic field in the quark determinant can
significantly change its expectation value. In the following, we explicitly
demonstrate this by calculating the dependence of $P$ on the external magnetic field in the continuum limit. To extrapolate the Polyakov loop to the limit $a\to0$, we perform its proper renormalization, considering the scheme suggested in ref.~\cite{Borsanyi:2012uq}, which we generalize to the case of nonzero magnetic fields. This renormalization is of the multiplicative form
\be
P_r(a,T,B) =  Z(a,T) \cdot P(a,T,B),
\label{eq:pren0}
\ee
and is discussed in detail in App.~\ref{app:renormP}.

\begin{figure}[t!]
\centering 
\includegraphics[width=.51\textwidth]{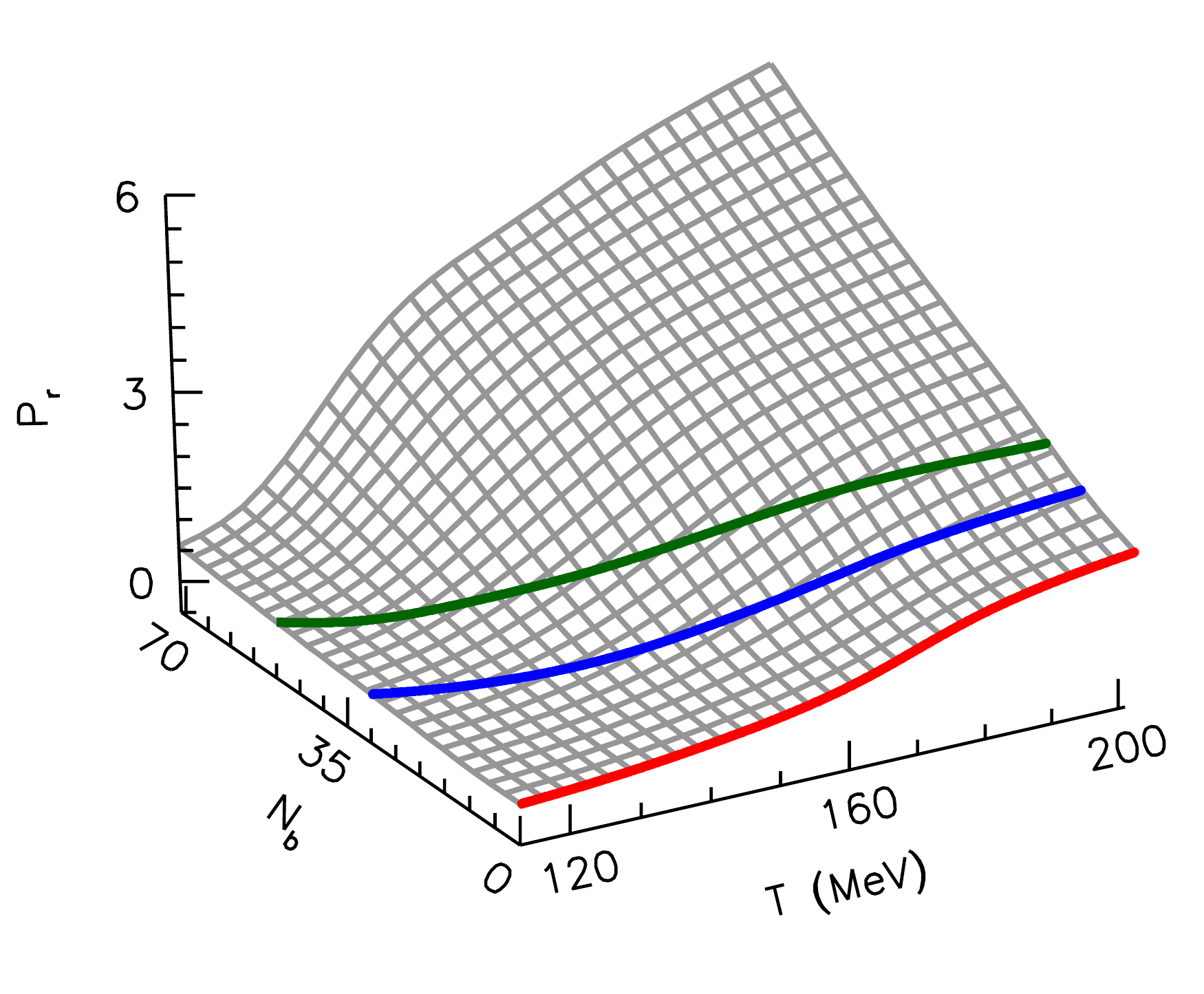}
\hspace*{-.2cm}
\includegraphics[width=.49\textwidth]{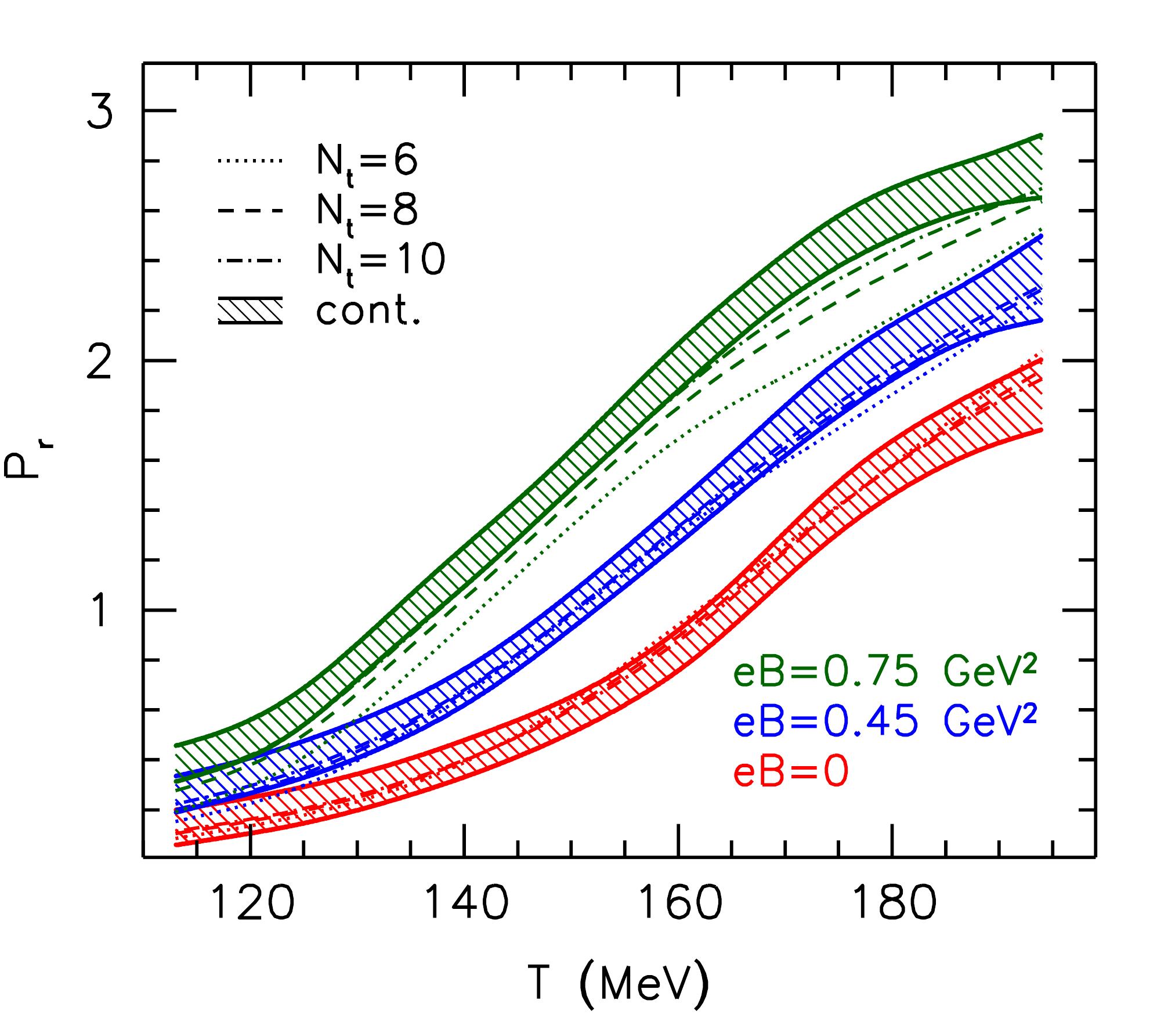}
\hspace*{-.2cm}
\caption{\label{fig:ploop_cont} Left panel: the temperature- and magnetic
  flux-dependence of the renormalized Polyakov loop in the continuum
  limit. The solid lines represent curves of constant magnetic field ($eB\sim N_b T^2$
  values as in the right panel). Right panel: the dependence of $P_r$ on the
  temperature around the crossover region. The different types of curves
  indicate lattice results obtained with different lattice spacings (different
  temporal extents). The shaded areas show the continuum extrapolations
  together with their uncertainty. }
\end{figure}

For the Polyakov loop calculation, we again employ the gauge configurations of ref.~\cite{Bali:2011qj}, generated with physical quark masses at various
values of the temperature and the magnetic flux $N_b\sim eB/T^2$. In any finite volume, 
this flux is quantized~\cite{'tHooft:1979uj}.
In order to determine the Polyakov loop as a function of
$T$, along a line of constant $eB$, an interpolation between the different
fluxes $N_b$ is necessary. We carry out this
interpolation in a systematic manner, by fitting our data points for all
temperatures, magnetic fluxes and lattice spacings altogether by a lattice
spacing-dependent, two-dimensional spline function. A similar spline fit is
described in ref.~\cite{Endrodi:2010ai}. Due to the scaling properties of the
action we use, the dependence on the lattice spacing is expected to be
quadratic. We incorporated this in the fit by having two parameters on each
node point as $p_1+p_2\cdot a^2$. Taking $eB=\textmd{const.}$ slices of this
two-dimensional surface at a certain $a$ gives the Polyakov loop for that
particular lattice spacing, while the $a=0$ surface corresponds to the
continuum limit.

In the left panel of Fig.~\ref{fig:ploop_cont}, we plot the continuum
extrapolated renormalized Polyakov loop $P_r$ as a function of the temperature and
the magnetic flux. The solid lines upon the surface correspond to $eB=0$,
$eB=0.45\textmd{ GeV}^2$ and $eB=0.75\textmd{ GeV}^2$ slices. In the right panel of
the figure, we show the temperature dependence of $P_r$ for these magnetic
fields on the three lattice spacings, together with the continuum
extrapolation. The shaded bands represent here the uncertainty of the continuum
extrapolated $P_r$. The results clearly show, that the Polyakov loop
\emph{increases} sharply with the magnetic field around $T_c$, and that this
feature persists in the continuum limit as well. As an empirical finding from that figure, inflection
points of these curves are not very precisely defined, but the transition
temperature from the renormalized Polyakov loops clearly decreases with the
magnetic field.

In the previous sections, we saw that low Dirac modes are the key to
understanding inverse magnetic catalysis. To complete the picture, we would
like to discuss one more point, namely the relationship between the low Dirac
modes and the Polyakov loop. It is well-known that light dynamical fermions
break the $Z(3)$ center symmetry of the quenched theory, by forcing the system
into the real Polyakov loop sector. This can be most easily understood
starting from a free field picture. If the gauge field background is trivial
-- apart from a spatially constant Polyakov loop -- the lowest Dirac
eigenmodes are constant in space, and change smoothly in the time direction to
fulfill the boundary condition. 
The boundary condition and the phase of the Polyakov loop
then combine to give the effective boundary condition (twist). 
The lowest fermion eigenvalue $\lambda_{\rm min}$ -- a generalized Matsubara frequency -- is
proportional to this temporal twist (see discussion around Eq.~(\ref{eq:evs_free})). For the usual antiperiodic boundary conditions, the twist is maximized when the Polyakov loop is in the real sector. Therefore, $\lambda_{\rm min}$ is also maximized for real Polyakov loop. Since the
difference in the quark action between the Polyakov loop sectors is dominated
by the small Dirac eigenvalues \cite{Kovacs:2008sc}, the real Polyakov loop
sector, the one with the largest $\lambda_{\rm min}$, will result in the
largest determinant.

It turns out that a qualitatively similar picture applies also to the
interacting theory above the transition temperature: a larger temporal twist
pushes the lowest modes higher up in the spectrum, away from
zero~\cite{Chandrasekharan:1995gt,Gattringer:2002tg,Bornyakov:2008bg}. As a result, the fermion determinant prefers the real Polyakov loop sector in the
$\mathrm{SU}(3)$ case. This is how the $\mathrm{Z}(3)$ symmetry is broken by
dynamical fermions. If the Dirac operator has a
tendency of having more small modes, the ordering of the Polyakov loop can
shift more small eigenvalues up, and the dependence of the determinant on the
Polyakov loop is stronger. This results in a stronger symmetry breaking effect
and a stronger ordering of the Polyakov loop.  This is exactly what
happens in the presence of the magnetic field.

We stress that an overall shift in the fermion action by the magnetic field would
not produce any physical effect. The important point is that this shift
fluctuates, and its fluctuations correlate with physical quantities (Polyakov
loop, condensate). This is what produces the reweighting effect that we
discussed. 
We have demonstrated this by virtue of scatter plots in
Figs.~\ref{fig:scatter1} and \ref{fig:scatter2}. For a qualitative
understanding, let us extend the Matsubara picture by magnetic fields. As the calculation 
in App.~\ref{app:free} shows, the free energies at nonzero $B$ favor
larger, i.e.\ deconfined Polyakov loops. Here, we focus on the main effect for
large $B$, where only the lowest Landau level is occupied. Its eigenvalue is independent\footnote{
 for spin 1/2 particles without anomalous magnetic moment}
of $B$, but
its degeneracy still increases proportional to the magnetic flux. The lowest
mode $\lambda_{\text{min}}$ of a given configuration thus acts as a ``handle''
for the magnetic field to change the fermionic determinant, to a first
approximation proportional to $(\lambda_{\text{min}})^{|qB|}$, suppressing
configurations with small eigenvalues even further. At the same time, configurations with many small eigenvalues are correlated to small Polyakov loops and large condensates, which are therefore suppressed as well.
This is what is observed in full QCD around the
transition. We repeat that the validity of this picture depends on low mode
dominance and the sensitivity of eigenmodes and condensates to Polyakov loops. Both are expected to hold for temperatures above the transition. It
would be interesting to study in detail how the structure of the low quark
modes change due to the magnetic field. An exciting clue in this direction
might be the presence of independent low modes at high temperatures, localized
to Polyakov loop islands \cite{Bruckmann:2011cc}. The magnetic field might
have a more coherent effect on these localized modes than on the bulk modes,
following random matrix statistics. However, a detailed study of this
mechanism is beyond the scope of the present paper.

We note that previously in the PNJL model~\cite{Gatto:2010pt} and also in
lattice simulations with larger-than-physical quark
masses~\cite{D'Elia:2010nq} the Polyakov loop was observed to decrease with
the magnetic field, and the transition temperature was shifted up by $B$. 
We conjecture that this is the reason why in those cases
inverse magnetic catalysis was not observed. Inverse magnetic catalysis
depends crucially on the interaction between the quark determinant and the
Polyakov loop, which happens through the lowest part of the Dirac spectrum. This
delicate effect has to be properly accounted for, to get the full physical
picture. Since the $\mathrm{Z}(3)$ symmetry breaking of the quark determinant depends 
strongly on the quark mass, lighter quarks are expected to enhance
inverse magnetic catalysis.
Let us finally remark that the latest results of the authors of ref.~\cite{D'Elia:2010nq} 
also indicate the sea condensate to decrease 
with $B$ for higher temperatures $T\sim 220 \textmd{ MeV}$, where the Polyakov loop 
is observed to increase with growing $B$~\cite{Deliatalk}, thus exhibiting the same 
mechanism as described above.

\section{Conclusions}
    \label{sec:C}

Using lattice simulations, we have analyzed how an external magnetic field
affects the QCD vacuum at finite temperatures.  Let us summarize: when the
magnetic field is switched on, the quark determinant suppresses gauge
configurations, on which the Dirac operator with the magnetic field has many
small eigenvalues. Around and above $T_c$, this suppression can happen by
ordering the Polyakov loop. This ordering effect is particularly efficient around $T_c$, where
the Polyakov loop effective potential is flat, and the magnetic field in the
determinant has a significant ordering effect on the Polyakov
loop. An ordered Polyakov loop implies that the lowest
Dirac eigenmodes become similar to Matsubara modes, and are shifted up in the
spectrum. For small quark mass, the quark condensate gets the largest
contribution from the low-end of the Dirac spectrum. Therefore, fewer low
eigenmodes also imply a suppression of the quark condensate. This
suppression of the condensate that we called ``sea effect'', competes with the
enhancement of the condensate due to the magnetic field in the operator
(``valence effect''). Due to the above described reasons, the sea suppression is
particularly efficient around $T_c$, and it actually overwhelms the valence
enhancement, resulting in inverse magnetic catalysis. The unifying theme in our
discussion is the lowest part of the Dirac spectrum. The response of low Dirac
modes to the magnetic field can explain the valence and sea effects, as well
as the ordering of the Polyakov loop.

The low mode dominance, which we have used in our arguments, and illustrated in
Fig.~\ref{fig:reweightfac}, strongly relies on the smallness of the quark
masses. One might speculate that the effect of inverse magnetic catalysis
becomes even more pronounced closer to the chiral limit. In any case, we believe to
have identified an important mechanism at work in the transition region,
already for magnetic fields slightly above the QCD scale (in contrast to
asymptotically large ones). It would be useful to incorporate such effects
back into low-energy effective models for QCD. Our results indicate that a
proper account of the interaction between the Polyakov loop and the Dirac
determinant and its influence on the condensate is needed for that.

\acknowledgments We thank the ECT$^\star$ in Trento for giving us the opportunity
to discuss some of the ideas presented in this paper with numerous
colleagues at the ``Workshop on QCD in strong magnetic fields'' in Trento,
Italy. 
Moreover, we thank Gunnar Bali, Szabolcs Bors\'anyi and S\'andor Katz for enlightening discussions, and 
Massimo D'Elia and Andreas Schmitt for useful correspondence.
T.G.K.\ is supported by the Hungarian Academy of Sciences under
``Lend\"ulet'' grant No.\ LP2011-011 and also partly by the EU Grant
(FP7/2007-2013)/ERC No. 208740. 
G.E.\ acknowledges support from the EU (ITN STRONGnet 238353) and F.B.\ from the DFG (BR 2872/4-2) as well as from the Alexander von Humboldt foundation.

\appendix
\section{Renormalization in the valence and sea sectors}
\label{app:renormseaval}

In this appendix, we show that the valence and sea condensates $\Delta\Sigma^{\rm val}$ and $\Delta\Sigma^{\rm sea}$ are properly renormalized, following the argumentation of refs.~\cite{Szabolcstalk,Szabolcscorrespondence}. Let us write down again the partition function, as in Eq.~(\ref{eq:partfunc}), for a single quark flavor with charge $q_1$,
\be
\Z \equiv \Z_{q_1B} = \int \D U\, e^{-S_g} \det \left(\slashed{D}(q_1B)+m\right),
\ee
from which the condensate is derived to be
\be
\bar\psi\psi(B) = \frac{T}{V}\frac{\partial \log\Z_{q_1B}}{\partial m} = \frac{T}{V}\expv{\Tr (\slashed{D}(q_1B)+m)^{-1}}_{q_1B},
\ee
where the subscript indicates that the expectation value is evaluated with respect to the partition function $\Z_{q_1B}$. 
The additive divergences of the condensate cancel in the change $\Delta$ with respect to the magnetic field. To eliminate the multiplicative divergence, a further multiplication by the quark mass is necessary, as in Eq.~(\ref{eq:pbpren}). Altogether, the combination $\Delta\Sigma\sim m \Delta\bar\psi\psi$ is ultraviolet finite.
On the other hand, in the ensemble described by $\Z_{q_1B}$, the change in the valence and sea condensates of Eq.~(\ref{eq:pbpvalsea}) read
\be
\begin{split}
\Delta\Sigma^{\rm val}\sim m\Delta\bar\psi\psi^{\rm val} = \frac{T}{V}\cdot m \left[ \expv{\Tr (\slashed{D}(q_1B)+m)^{-1}}_{0} - \expv{\Tr (\slashed{D}(0)+m)^{-1}}_{0}\right], \\
\Delta\Sigma^{\rm sea}\sim m\Delta\bar\psi\psi^{\rm sea} = \frac{T}{V}\cdot m \left[ \expv{\Tr (\slashed{D}(0)+m)^{-1}}_{q_1B} - \expv{\Tr (\slashed{D}(0)+m)^{-1}}_{0}\right], \\
\end{split}
\label{eq:appvalsea}
\ee
Our objective is to show that these combinations are also properly renormalized.

To this end, let us consider an additional quark flavor, with mass $m_2$, charge $q_2$, and multiplicity $\epsilon$. Even though the multiplicity is in practice restricted to take integer values, in principle one can also define a theory with a real multiplicity $\epsilon\in\mathds{R}$, described by the partition function,
\be
\Z_{q_1B+\epsilon q_2 B} = \int \D U\, e^{-S_g} \det \left(\slashed{D}(q_1B)+m\right) \det \left(\slashed{D}(q_2 B)+m_2\right)^{\epsilon}.
\ee
The condensate of the additional quark flavor is correspondingly given by
\be
\bar\psi_\epsilon\psi_\epsilon(B)
= \frac{T}{V}\frac{\partial \log\Z_{q_1B+\epsilon q_2 B}}{\partial m_2}
= \frac{T}{V}\cdot \epsilon\cdot \expv{\Tr (\slashed{D}(q_2 B)+m_2)^{-1}}_{q_1B+\epsilon q_2 B},
\label{eq:psibarpsi_0}
\ee
where this time the expectation value is evaluated in the ensemble with both the original and the additional flavor taken into account. 
Let us now expand Eq.~(\ref{eq:psibarpsi_0}) around $\epsilon=0$. Since $\det^\epsilon = 1+\mathcal{O}(\epsilon)$, we get the expectation value in the $\Z_{q_1B}$ ensemble plus higher order corrections,
\be
\bar\psi_\epsilon\psi_\epsilon(B)
= \frac{T}{V}\cdot \epsilon\cdot \expv{\Tr (\slashed{D}(q_2 B)+m_2)^{-1}}_{q_1B} + \mathcal{O}(\epsilon^2).
\ee 
Again, the combination $m_2 \Delta \bar\psi_\epsilon\psi_\epsilon$ is completely renormalized for any value of $\epsilon$. 
Thus, in its expansion with respect to $\epsilon$, all terms must also separately be ultraviolet finite. 
In particular, the $\mathcal{O}(\epsilon)$ term at $m_2=m$
\be
\frac{T}{V} \cdot m\left[ \expv{\Tr (\slashed{D}(q_2 B)+m)^{-1}}_{q_1B} - \expv{\Tr (\slashed{D}(0)+m)^{-1}}_{0}\right]
\label{eq:q1q2}
\ee
is properly renormalized, for any values of the charges $q_1$ and $q_2$.

Upon substituting $q_1=0$ and renaming $q_2=q_1$, Eq.~(\ref{eq:q1q2}) exactly equals the valence condensate of Eq.~(\ref{eq:appvalsea}). On the other hand, substituting $q_2=0$ gives the sea condensate. Altogether, this shows that the combinations $\Delta\Sigma^{\rm val}$ and $\Delta\Sigma^{\rm sea}$, as defined after Eq.~(\ref{eq:pbpren}), are properly renormalized. The above argumentation relies essentially on the fact that the magnetic field enters only in the combination $qB$. Therefore, varying the charges effectively allows for varying the magnetic field.

\section{Free case calculation}
\label{app:free}

We consider the free energy of (one flavor of) massive quarks in the continuum at finite temperature and study the interplay of constant Polyakov loops, mass and magnetic field (and chemical potential as a cross-check).

Let the gauge background be an abelian field generating a constant magnetic field $(0,0,B)$, e.g.\ $a_x=0,\,a_y=B x$, plus a constant SU(3) gauge field $A_0$ generating a diagonal and real Polyakov loop,
\be
L=\left(\begin{array}{ccc}e^{2\pi i \p}&&\\&e^{-2\pi i \p}&\\&&1
  \end{array}\right)\,,\qquad 
\frac13\,\Tr\, L=\frac13\,\big(1+2\cos(2\pi\p)\big).
\ee
Obviously, $\p=1/3$ amounts to a traceless Polyakov loop, which we will take as a signal for the confined phase, while for $\p\to 0$ the Polyakov loop becomes trivial, $L\to \mathds{1}$, which stands for the deconfined phase.

In the computation of the eigenvalues of the Dirac operator $\slashed{D}=\gamma_\mu D_\mu$ with covariant derivative $D_\mu=\partial_\mu+iA_\mu+iqa_\mu$ at finite temperature $T$, the Polyakov loop background can be incorporated in the quark boundary conditions in each color sector as $\psi_a(x_0+1/T,\vec{x})=-\exp(2\pi i \chi_a)\psi_a(x_0,\vec{x})$ with the three phases $\chi_a\in\{\p,-\p,0\}$. This modifies the Matsubara frequencies to $(\pi +2\pi\chi_a+2\pi k)T$ with integer $k$.

The eigenvalues of the squared Dirac operator in each color sector read
\be
-\slashed{D}^2_a(B)\to
\lambda^2_a(B)=\left(2\pi(1/2+\chi_a+k)T\right)^2
+p_z^2+|qB|(2n+1)+qBs,
\label{eq:evs_free}
\ee
with momenta $p_z\in\mathds{R}$, Landau levels $n=0,1,\ldots$ and $s=\pm 1$ being twice the spin. The degeneracy of these eigenvalues is proportional to the magnetic flux $B\cdot A$ with $A$ the area of the system perpendicular to the magnetic field. Together with the $z$-extension $L_z$ in the $p_z$ integration, this yields the three-volume $V=AL_z$ in
\be
\sum_{\lambda^2}=2\frac{V}{(2\pi)^2}\,|qB|\sum_{k=-\infty}^\infty\int\!\!dp_z\sum_{n=0}^\infty\, \sum_{s=\pm 1}.
\label{eq:lambdas}
\ee
The factor 2 comes from the degeneracy of particles and antiparticles at given spin (both together provide the four eigenvalues of the four-dimensional gamma matrices).

For the partition function we make use of chiral symmetry to write
\be
\log Z(B)=\sum_{a=1}^3\log \det(\slashed{D}_a(B)+m)=\frac12\sum_a\sum_{\lambda^2}\log (\lambda_a^2(B)+m^2).
\label{eq app log Z}
\ee
The proper time/zeta function regularization formalism can be introduced via the Mellin transform
\be
\log y =-\left.\frac{\partial}{\partial\a}\right|_{\a \to 0} y^{-\a}
 =-\left.\frac{\partial}{\partial\a}\right|_{\a \to 0} \frac1{\Gamma(\a)}
  \int_0^\infty\!\!d\s\, \s^{\a-1}\exp(-\s y),
\ee
(valid for Re $y$, Re $\alpha>0$). 
Through interchanging the order of limits, integrals etc.\ analytic continuation is performed and  (some of) the singularities are removed. This technique has been used in the classical papers  \cite{Heisenberg:1935qt,Schwinger:1951nm}, for a review see e.g.\ \cite{Dunne:2004nc}.

Using Eqs.~(\ref{eq:lambdas}-\ref{eq app log Z}), the free energy density $f=-T\log Z/V$ for a given background becomes
\be
f(\p; B)=\frac{qB}{2\pi^2}\sum_a \int_0^\infty\!\!d\s\,\s^{-2}\exp(-m^2 \s)\coth(qB\s)\,
  \theta_3\left((\chi_a+1/2)\pi,e^{-\frac1{4 s T^2}}\right)\,,
\ee
where we have assumed positive $qB$ for simplicity and performed the Gaussian integration over $p_z$, the derivative wrt.\ $\a$ and the sums over $n$, $s$ and $k$. The latter introduced an elliptic theta function $\theta_3(u,q)\equiv1+2\sum_{n=1}^\infty q^{n^2}\cos(2nu)$.

The $s$-integral is known to be divergent for small $s$ (as $\coth(qB\s)\to 1/(qB\s)$ and 
$\theta_3\big(\ldots,\exp(-\frac1{4 s T^2})\big)\to 1$). Since $s$ has dimension length squared, this is a UV-divergence. After subtracting the $B=0$ part it is weakened to be logarithmic and can be absorbed in the renormalization of the electric charge~\cite{Schwinger:1951nm}.

Here we follow a different route and consider the difference of the free energy of a given background $\p$ and the deconfining background $\p=0$, which means
\begin{align}
&\Delta_\p f(B) \equiv f(\p; B)-f(0; B)\label{eq:delta_f_B}\\
 &=\frac{qB}{\pi^2}\int_0^\infty\!\!d\s\,\s^{-2}\exp(-m^2 \s)\coth(qB\s)\,
  \left[\theta_3\left((\p+1/2)\pi,e^{-\frac1{4 s T^2}}\right)- 
   \theta_3\left(\pi/2,e^{-\frac1{4 s T^2}}\right)\right]\nonumber
\end{align}
where we used the $\p\to -\p$ symmetry. This integral is finite because the nonperturbative behavior $\exp(-\#/\s)$ of the theta function difference for small $s$ now kills all power-like singularities.

Note, that the integrand in (\ref{eq:delta_f_B}) is positive definite and so is $\Delta_\p f$. Hence the quark determinant always favors $\p=0$, i.e. deconfining Polyakov loops.  Furthermore, the integrand increases monotonically as $m$ is reduced, such that this effect becomes more pronounced for light quarks. This is consistent with the finding from `lattice experiments' that the onset of the deconfined phase -- the deconfinement transition -- occurs at lower temperatures, when the quarks become lighter. 

As another cross-check we consider the effect of an imaginary chemical potential $\mu=i\eta$. It can also be incorporated in the boundary condition (for all color sectors in the same way) and the corresponding change of free energies follows from Eq.~(\ref{eq:delta_f_B}) in the $B\to 0$ limit is
\be
\begin{split}
 \Delta_\p f(\eta) &\equiv f(\p; \eta)-f(0; \eta) \\
 &=\frac{1}{2\pi^2}\int_0^\infty\!\!d\s\,\s^{-3}\exp(-m^2 \s)
    \left[\theta_3\left((\p+1/2-\eta)\pi,e^{-\frac1{4 s T^2}}\right)\right.\\
  &\quad\left.+\,\theta_3\left((-\p+1/2-\eta)\pi,e^{-\frac1{4 s T^2}}\right)
      -2\,\theta_3\left((1/2-\eta)\pi,e^{-\frac1{4 s T^2}}\right)\right]
\end{split} 
\ee
Again, the integrand is positive definite, and the effect of the quark mass is the same. Increasing $\eta$ we have observed $\Delta_\p f(\eta)$ to decrease, although the integrand does not change monotonically with $\eta$\footnote{The first derivatives of $\Delta_\p f(\eta)$ with respect to $\eta$ and $\p$ vanish at $\eta=\p=0$, i.e.\ to linear order in the imaginary chemical potential and around the deconfined phase. The second derivatives with respect to $\eta$ and $\p$ at $\eta=\p=0$ give $T^2$ with a negative coefficient, i.e.\ the decrease mentioned.}. Thus an imaginary chemical potential $\eta$ favors $\p>0$, i.e. confining Polyakov loops, consistent with the finding that the transition temperature grows with it.

After these checks, we analyze the effect of an external magnetic field in the same fashion. For this the dependence of the integrand in Eq.~(\ref{eq:delta_f_B}) on $qB$ is decisive: the function $qB\cdot\coth(qB\s)$ increases with $qB$ for all $\s$. Hence a \emph{magnetic field favors deconfined Polyakov loops} and tends to lower the transition temperature. This is certainly a sea effect as it comes from the quark determinant in the effective action, and it gets washed out by large masses $m$. 
This effect is already present in the lowest Landau level ($n=0,\,s=-1$ in Eq.~(\ref{eq:evs_free})). The result of this projection is to replace $\coth(qBs)\to1$ in Eq.~(\ref{eq:delta_f_B}).
The integrand still increases proportional to $qB$, from the degeneracy of the modes.

It is interesting to note that these findings are inverted when modifying either the statistics, the boundary condition or the spin of the quarks: for commuting $c$-number quarks the determinant has to be replaced by its inverse giving a minus sign to the free energy. For quarks periodic in the time-like direction one has to remove $\pi/2$ in both $\theta_3$'s and the integrand becomes negative definite. For spinless quarks $qB\cdot\coth(qB\s)$ gets replaced by $qB/\sinh(qB\s)$ which decreases with $qB$.

We conclude this section by a brief discussion on the validity of these considerations. First of all, the approximation of QCD by free quarks is expected to work at high temperatures thanks to asymptotic freedom. The free energy, for instance, approaches the Stefan-Boltzmann result and the spectral gap has been found to be described by the lowest eigenvalue in the free case \cite{Gavai:2008xe,Bruckmann:2011cc} (with background Polyakov loop as we do here).

The (de)confinement order parameter is the spatially averaged Polyakov loop. It is this quantity that changes from zero to non-zero (small to large in a crossover) as the temperature rises. The local Polyakov loops strongly vary, as can be seen from the decaying Polyakov loop correlator (related to one definition of confinement) and the distribution of local Polyakov loops (which in the low temperature phase is known to be the Haar measure). Such a local resolution of the Polyakov loop is lost in our approach as we compare spatially constant backgrounds. Thus, we cannot speak about transitions between disordered and ordered Polyakov loops.  Nevertheless, the finding that free quarks favor constant trivial Polyakov loop backgrounds in comparison to constant nontrivial ones can be interpreted as a signal for the suppression of the confined phase by quarks in realistic QCD.
This effect has the right tendency under a change of the control parameters quark mass and  (imaginary) chemical potential. Moreover, as we have demonstrated, for external magnetic fields it correctly describes the increase of the Polyakov loops and the decrease of the transition temperature as seen in full QCD.

\section{Renormalization of the Polyakov loop}
\label{app:renormP}

In this appendix we discuss the renormalization of the Polyakov loop $P$. 
First of all, we note that $P$ is related to the free energy of an infinitely separated, static quark-antiquark pair $F_{\bar q q}=-2T\log P$, which contains additive divergences in the cutoff. These additive divergences, therefore, appear as multiplicative divergences in $P$. A possible scheme for the corresponding renormalization is to set the renormalized free energy to a fixed value $F_\star$, at $B=0$ and a fixed but arbitrary temperature $T_\star>T_c$. Since the magnetic field introduces no new divergences in $F_{\bar qq}$ (see below), this subtraction is unchanged at $B>0$, making the combination
\be
F_{\bar q q,r}(T,B) = F_{\bar q q}(T,B)-F_{\bar qq}(T_\star,B=0)+ F_\star
\label{eq:Fqqrenorm}
\ee
finite in the continuum limit.

To see that the divergences of $F_{\bar qq}$ are indeed $B$-independent, let us consider the {\it total} free energy $F$ of the system, in the presence of an external magnetic field. $F$ includes both the contribution of matter, and that of the field itself, $B^2/2$. It has been known for long~\cite{Schwinger:1951nm} that the only $B$-dependent divergence, which resides in $F(B)$, is eliminated via the simultaneous renormalization of the electric charge $q$, and that of the pure magnetic term $B^2/2$ (see also refs.~\cite{Dunne:2004nc,Endrodi:2013cs}), which leaves the combination $qB$ invariant. On the other hand, the free energy $F_{\bar qq}$ of the static color charges only couples to the magnetic field via sea quark loops, in the form $qB$, and contains no pure magnetic term. Then, it also cannot contain any $B$-dependent divergence.

One can also explicitly show the absence of $B$-dependent divergences in $F_{\bar qq}$ using perturbation theory. Namely, the leading order perturbative term in $F_{\bar qq}$ is given by the exchange of a gluon. The latter creates a virtual (sea) quark loop, to which $B$ couples. This diagram is the same as the quark loop with two photon lines, up to constant group theoretical factors. Using the exact quark propagator in the magnetic field~\cite{Schwinger:1951nm}, this photon self-energy has been explicitly calculated at asymptotically large magnetic fields to be finite~\cite{Fukushima:2011nu}. Diagrams corresponding to higher orders in perturbation theory contain the same diagram as subdiagram, or sea quark loops with even more gluon lines attached, which (since they contain a higher number of quark propagators) are also finite. Altogether this shows again that the renormalization of $F_{\bar q q}$ -- and, accordingly, that of $P$ -- is independent of $B$.

We can now translate the renormalization condition~(\ref{eq:Fqqrenorm}) for $F_{\bar qq}$ to the renormalization of the Polyakov loop.  
On a lattice with spacing $a$, the corresponding renormalization for $P=\exp(-F_{\bar qq}/2T)$ amounts to
\be
P_r(a,T,B) =  Z(a,T) \cdot P(a,T,B),
\quad\quad\quad
Z(a,T) = \left(\frac{P_\star}{P(a,T_\star,B=0)} \right)^{T_\star / T},
\label{eq:pren}
\ee
where $P_\star=\exp(-F_\star/2T_\star)$. 
We choose $T_\star = 162 \textmd{ MeV}$, where the bare Polyakov loops are already significantly nonzero, and we set $F_\star=0$ ($P_\star=1$). Different values for these 
parameters correspond to different renormalization schemes, connected by finite renormalizations.

\bibliographystyle{JHEP}
\bibliography{bfield}

\end{document}